\documentclass[aps,prl,reprint,superscriptaddress,floatfix]{revtex4-2}

\usepackage{graphicx}
\usepackage{dcolumn}
\usepackage{bm}
\usepackage{multirow}
\usepackage{hyperref}
\usepackage{amssymb,amsmath}
\usepackage{graphics,color}
\usepackage{subfigure,epsfig}
\usepackage{comment}
\usepackage{lipsum}
\usepackage{footnote}
\usepackage{booktabs,siunitx}
\usepackage[normalem]{ulem}
\newcommand{\tabhead}

\graphicspath{{images/}}

\begin{document}
\title{Intrinsic vibrational angular momentum from non-adiabatic effects \texorpdfstring{\\ in non-collinear magnetic molecules}{}}
\author{Oliviero Bistoni}
\affiliation{Sorbonne Universit\'e, CNRS, Institut des Nanosciences de Paris, UMR 7588, F-75252, Paris, France}
\affiliation{Dipartimento di Fisica, Universit\`a di Trento, Via Sommarive 14, 38123 Povo, Italy}
\author{Francesco Mauri}
\affiliation{Dipartimento di Fisica, Universit\`a di Roma La Sapienza, Piazzale A. Moro 5, I-00185 Roma, Italy}
\author{Matteo Calandra}
\affiliation{Sorbonne Universit\'e, CNRS, Institut des Nanosciences de Paris, UMR 7588, F-75252, Paris, France}
\affiliation{Dipartimento di Fisica, Universit\`a di Trento, Via Sommarive 14, 38123 Povo, Italy}

\begin{abstract}
We show that in non-collinear magnetic molecules, non-adiabatic (dynamical) effects due to the electron-vibron coupling are time-reversal symmetry breaking interactions for the vibrational field. As in these systems the electronic wavefunction can not be chosen as real, a nonzero geometric vector potential (Berry connection) arises. As a result, an intrinsic nonzero  vibrational angular momentum occurs even for non-degenerate modes and in the absence of external probes. The vibronic modes can then be seen as elementary quantum particles carrying a sizeable angular momentum.
As a proof of concept, we demonstrate the magnitude of this topological effect by performing non-adiabatic first principles calculations on platinum clusters and by showing that these molecules host sizeable intrinsic phonon angular momenta comparable to the orbital electronic ones in itinerant ferromagnets.
\end{abstract}

\maketitle

Several experiments demonstrated the non-negligible interaction between vibrational modes and magnetic fields or optical probes. The phonon Hall effect~\cite{PhysRevLett.95.155901,*Inyushkin2007,Zhang_2016} and the phonon contribution to the gyromagnetic ratio detected in the Einstein de Haas effect~\cite{EinsteinHaas1,*EinsteinHaas2,PhysRevLett.112.085503} are eminent examples.
Moreover, it has been demonstrated that valley selective infrared optical absorption in transition metal dichalcogenides  breaks time-reversal symmetry for the phonon field and can be used to probe the chirality of phonon modes at particular points in the Brillouin zone~\cite{Zhu579,PhysRevLett.115.115502}.  

In the absence of external probes, phonons are usually understood in terms of springs and as such they are considered as linearly polarized, so they do not break time-reversal symmetry and they are not supposed to carry a finite angular momentum.
An intrinsic phonon angular momentum can be obtained from a twofold degenerate vibrational mode, as a linear combination of two linear phonon eigenvectors can lead to a circularly polarized mode, in the same way as circularly polarized light can arise from two linear polarizations.
This case has been investigated in literature extensively  \cite{PhysRevLett.115.115502,PhysRevLett.119.255901,Xu_2018,PhysRevB.98.134304,PhysRevB.100.094303}, particularly for hexagonal crystal lattices \footnotemark[4].
\footnotetext[4]{In hexagonal crystal lattices with broken in-plane inversion symmetry (i.e. BN monolayer), single degenerate phonon modes carry opposite angular momenta at the Brillouin zone corners ${\bf K}$ and ${\bf K^{\prime}=-{\bf K}}$. However, the phonon modes at ${\bf K}$ and $\bf K^{\prime}$ have the same energy and if vibrations are described in a  $\sqrt{3}\times\sqrt{3}R30^{o}$ supercell, ${\bf K}$ and $\bf K^{\prime}$ fold at $\bf\Gamma$ and the two single degenerate modes become twofold degenerate at zone center.}
For each circularly polarized phonon carrying an angular momentum $\bm{\ell}$ there exists another linearly independent combination of linear polarizations leading to an angular momentum $-\bm{\ell}$, so that the total phonon angular momentum for the degenerate modes is zero. An external time-reversal symmetry breaking probe, such as optical absorption or an external magnetic field is then needed to break the degeneracy.

The question is still open if a non-degenerate phonon mode can host an intrinsic angular momentum without external probes. Namely, can an intrinsic mechanism lead to a time-reversal symmetry breaking in the phonon field?

In this work we demonstrate that non-adiabatic (dynamical) effects due to the  electron-vibron interaction generate synthetic gauge fields in insulating non-collinear magnetic molecules.
We provide the microscopic link between topology and the electron-vibron interaction by showing that in these systems a nonzero Berry curvature leads to a finite intrinsic vibrational angular momentum even for non-degenerate modes and in the absence of external magnetic fields. 
As a proof of concept, we demonstrate the effect by performing non-adiabatic first principles calculations on platinum clusters.

We introduce a cumulative index $\lambda=(I,\alpha)$ for the Cartesian coordinates $\alpha=x,y,z$ of the $I^{\rm th}$ atom in a molecule. The atomic position in a molecule is $R_{\lambda}=R_{\lambda}^\text{eq}+ u_{\lambda}$ where $R_\lambda^\text{eq}$ are the coordinates of the atomic equilibrium positions and $u_\lambda$ is the Cartesian component of the ionic displacement of the $I^{\rm th}$ atom. 
In the Born-Oppenheimer approximation the quantum-mechanical Hamiltonian for the ionic motion reads \cite{PhysRevB.86.104305}
\begin{equation}    \label{Eq:Hamiltonian}
    \mathcal{H}=\frac{1}{2M}\sum_\lambda\left[p_\lambda-\hbar{\mathcal{A}}_{\lambda}(\bm u)\right]^2+ E(\bm u)
\end{equation}
where  $p_\lambda=-i\hbar \nabla_{u_\lambda}\equiv-i\hbar \nabla_\lambda$ is the ionic momentum, $\mathcal{A}_{\lambda} (\bm u)=i\langle\Psi(\bm u)|\nabla_{\lambda}\Psi(\bm u)\rangle$ is a geometric vector potential (the so-called Berry potential or Berry connection), $E(\bm u)$ is the potential energy felt by the ions due to the electrons and
$|\Psi({\bf u})\rangle$ is the ground-state electronic wavefunction which depends parametrically on the nuclear displacements $\bm u$.
For ease of notation, we consider equal masses for all the atoms, as this also corresponds to the case treated in this work.

In the absence of external magnetic fields and non-collinear magnetic order, the electronic wavefunction $|\Psi({\bf u})\rangle$ can be taken as real and the geometric vector potential in Eq.~\ref{Eq:Hamiltonian} is zero \cite{Resta2000}.
On the other hand, in the absence of an external magnetic field but for  a non-collinear magnetic molecule, the electronic wavefunction is complex and cannot be chosen as real, so that $\mathcal{A}_\lambda(\bm u)$ is nonzero and nontrivial geometric effects may occur.

The Berry curvature is defined as
$\Omega_{\lambda\eta}=\partial_\lambda \mathcal{A}_{\eta}- \partial_\eta \mathcal{A}_\lambda$
where $\partial_\lambda\equiv\partial/\partial u_\lambda$.
In linear response theory $\Omega_{\lambda\eta}$ is independent on the parameter $\bm u$. 
In the Heisenberg representation, the equation of motion for the nuclear displacement reads \cite{PhysRevLett.123.255901}
\begin{equation}    \label{Eq:motion_time}
    M\ddot u_\lambda+ \hbar \sum_{\eta} \Omega_{\lambda\eta} \dot u_\eta +\partial_\lambda E= 0
\end{equation}
In the harmonic approximation we expand the potential energy up to the second order in the ionic displacement.
Using monochromatic solutions in $\omega$, the equation of motion can thus be written as
\begin{equation}    \label{Eq:motion_freq}
    \frac{1}{M} \sum_\eta \left[C_{\lambda\eta}-i\hbar\omega\Omega_{\lambda\eta}\right]e_\eta=\omega^2e_\lambda
\end{equation}
where $C_{\lambda\eta}=\partial_\lambda\partial_\eta E$ is the static harmonic force-constant matrix and $e_\lambda$ are the vibrational polarization vectors. The term in $\Omega_{\lambda\eta}$ is an effective Lorentz force exerted by the geometrical magnetic field (note that $\Omega_{\lambda\eta}$ is a real antisymmetric matrix).
Formal solution of the non-linear eigenvalue equation~\ref{Eq:motion_freq} can be found in the supplemental material of Ref.~\cite{PhysRevLett.112.085503}.

At zero temperature, the expectation value of the quantum vibrational angular momentum ${\bf L}=\sum_{I}{\bf u}_I\times\dot{\bf u}_I$ over the quantum vibron ground state reads \cite{PhysRevLett.112.085503} $\langle {\bf L}\rangle = \sum_{\nu}\bm{\ell}_\nu$, where $\nu$ labels the vibrational modes and $\bm{\ell}_\nu$ can be expressed in terms of (the Cartesian components of) the vibrational polarization vectors $\bm e_{I\nu}$ as
\begin{equation} \label{Eq:mode_angular}
    \bm{\ell}_\nu=-i\hbar\sum_I \bm e_{I\nu}^\ast\times\bm e_{I\nu}.
\end{equation}
We underline that the expectation value of the cartesian components of ${\bf L}$ over the vibron ground state, i.e.  $\langle L_{\alpha} \rangle = \sum_{\nu}\ell_{\nu,\alpha}$ for $\alpha=x,y,z$, are not quantized quantities and can assume any value. Indeed, while in the absence of non-collinear magnetism and in the presence of an external magnetic field
along the z-direction $L_z$ commutes with ${\cal H}$ in Eq. \ref{Eq:Hamiltonian}, in the case of non collinear magnetism treated here, the components of the phonon angular momentum $L_{\alpha}$ do not commute with ${\cal H}$. Thus, in our case the ground state of Eq. \ref{Eq:Hamiltonian} is not an eigenstate of $L_{\alpha}$ and neither $\sum_{\nu}\ell_{\nu,\alpha}$ nor $\ell_{\nu,\alpha}$ are quantized.

If the Berry curvature $\Omega_{\lambda\eta}$ in Eq.~\ref{Eq:motion_freq} vanishes (i.e. for real wavefunctions), the polarization vectors are eigenfunctions of the static and real force constant matrix $C_{\lambda\eta}$ and therefore they themselves are real (up to an irrelevant global phase factor) and the angular momentum $\bm \ell_\nu$ is equal to zero (since $\bm e^\ast=\bm e$).
On the contrary, in molecules with non-collinear magnetism,  the electronic wavefunctions are necessarily complex and the Berry curvature does not vanish. The polarization vectors of Eq.~\ref{Eq:motion_freq} are  therefore  intrinsically complex and  give rise to a nonzero vibrational angular momentum. Thus, the occurrence of a vibrational mode with a finite angular momentum is intimately connected with the existence of a geometrical or physical gauge field.

We now show that, in the independent electron approximation, both Eq.~\ref{Eq:motion_freq} and the existence of a nonzero intrinsic angular momentum for a molecular vibrational mode naturally arise from the theory of non-adiabatic (dynamical) effects developed in Ref.~\cite{PhysRevB.82.165111}, providing the link between the electron-vibron interaction, topological effects and the existence of a nonzero intrinsic  vibrational angular momentum.
Within time dependent density functional theory and in the adiabatic local density approximation, the dynamical force constant matrix for frequencies $\omega$  smaller than the HOMO-LUMO gap $(\Delta)$ reads:
\begin{equation}    \label{Eq:force_constants}
    C_{\lambda\eta}(\omega) = C_{\lambda\eta} + \Pi_{\lambda\eta}(\omega)
\end{equation}
where $C_{\lambda\eta}$ is the static force constant matrix
and $\Pi_{\lambda\eta}(\omega)$ can be written in perturbation theory as
\begin{multline}    \label{Eq:self-energy}
    \Pi_{\lambda\eta}(\omega)= 2 \sum_{m,n} \left[ \frac{f_{m}-f_{n}}{\epsilon_{m}-\epsilon_{n}+ \hbar\omega} -\frac{f_{m}-f_{n}}{\epsilon_{m}-\epsilon_{n}} \right] \times \\
    \times \langle \psi_{n}| \partial_\lambda H_{\rm KS} | \psi_{m}\rangle
    \langle \psi_{m}| \partial_\eta H_{\rm KS}| \psi_{n}\rangle.
\end{multline}
Here  $|\psi_{m}\rangle$, $\epsilon_m$ and $f_m$ are the Kohn-Sham  wavefunctions,  energy levels
and Fermi occupations 
at equilibrium positions (i.e. ${\bf u}=0$), respectively, and 
$H_{\rm KS}$ is the electronic Kohn-Sham hamiltonian. The deformation potential matrix element $\langle \psi_{m}| \partial_\eta H_{\rm KS}| \psi_{n}\rangle$ is related to the electron-vibron interaction.
The non-adiabatic (dynamical) vibrational frequencies ($\tilde\omega_\nu$) and polarization vectors ($\tilde e_{\eta\nu}$), which will be marked hereinafter with a tilde, are obtained from the non-linear eigenvalue equation
\begin{equation} \label{Eq:secular_equation}
\frac{1}{M}\sum_{\eta} C_{\lambda\eta}(\tilde\omega_\nu)\tilde e_{\eta\nu}= \tilde\omega_\nu^2 \tilde e_{\lambda\nu}
\end{equation}
In the case when $\hbar\omega\ll \Delta$, Eq. \ref{Eq:force_constants} can be expanded at first order to obtain
\begin{equation}    \label{Eq:force_const_Tay}
  C_{\lambda\eta}(\omega)= C_{\lambda\eta} -i\hbar\omega\Omega_{\lambda\eta}^{\rm KS}+{\cal O}(\omega^2)
\end{equation}
where $\Omega_{\lambda\eta}^{\rm KS}=\sum_m f_m\Omega_{\lambda\eta,m}^{\rm KS}$ and
$\Omega_{\lambda\eta,m}^{\rm KS}$ is
the Berry curvature of the $m^{\rm th}$ Kohn-Sham state~\cite{Resta2000} with respect to the atomic displacement, namely:
\begin{multline}    \label{Eq:curvature}
    \Omega_{\lambda\eta,m}^{\rm KS}= -2\ \text{Im} \sum_{n\neq m} \frac{1}{(\epsilon_{m}-\epsilon_{n})^2} \times \\
    \times \langle\psi_{m} |\partial_\lambda H_{\rm KS}|\psi_{n}\rangle
    \langle \psi_{n}|\partial_\eta H_{\rm KS}|\psi_{m}\rangle.
\end{multline}
The matrix $\Omega_{\lambda\eta}^{\rm KS}$ is a real antisimmetric
matrix that plays the role of $\Omega_{\lambda\eta}$ in the case of Kohn-Sham independent electrons.
The non-adiabatic (dynamical) vibrational frequencies $\tilde\omega_\nu$ and  polarization vectors $\tilde{\bf e}_{I\nu}$ can then be obtained as solutions of the non-linear eigenvalue equation~\ref{Eq:secular_equation} and used to calculate the quantum angular momentum via Eq.~\ref{Eq:mode_angular}.

The Eqs.~\ref{Eq:force_const_Tay} and \ref{Eq:curvature} are the microscopic link between the electron-vibron interaction, non-adiabatic (dynamical) effects and the occurrence of a finite angular momentum in molecules.
Furthermore, they provide a practical computational scheme of the vibrational quantum angular momentum using the theory proposed in Ref.~\cite{PhysRevB.82.165111}.
Since $\Omega_{\lambda\eta}$ is proportional to the square of the deformation potential and inversely proportional to the HOMO-LUMO gap, Eq.~\ref{Eq:curvature} suggests that large non-adiabatic (dynamical) effects and vibrational angular momenta could be found in non-collinear magnetic molecules with a small gap and a large electron-vibron interaction.

We demonstrate the occurrence of an intrinsic total vibrational
angular momentum due to non-adiabatic (dynamical) effects by
considering platinum clusters, namely a trimer Pt$_3$ and  a pentamer Pt$_5$. These systems are ideal
as they are (i) magnetic, (ii) the large spin-orbit coupling leads to non-collinear magnetic structures and
(iii) the HOMO-LUMO gap is very small.

\begin{figure}[htp]
\centering
\includegraphics[scale=0.4,angle=90,origin=t]{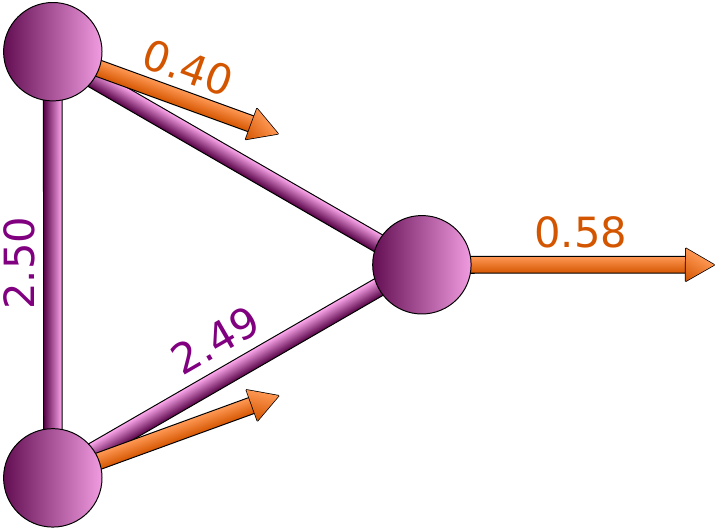}
\hspace{1 cm}
\includegraphics[scale=0.4]{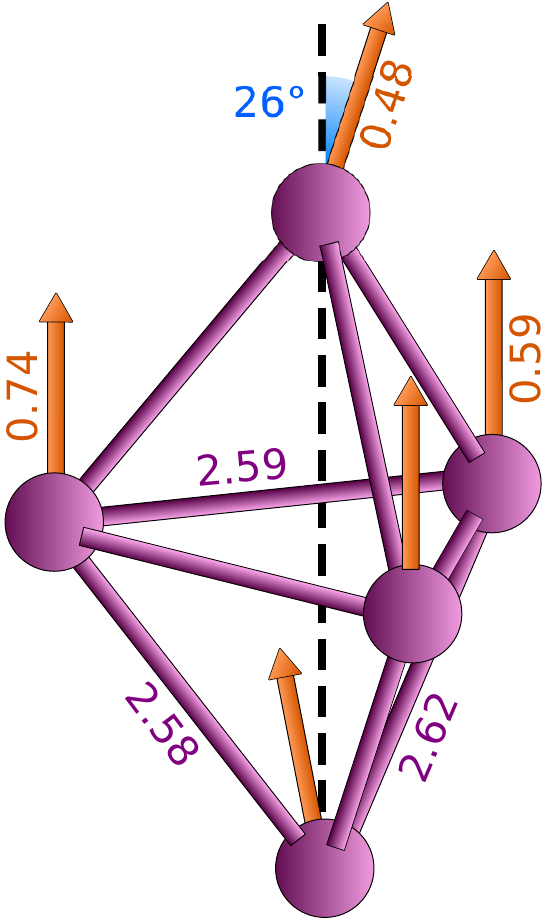}
\caption{Non-collinear magnetic ground state of Pt$_3$ and Pt$_5$: structure, interatomic distances and magnetic momenta.}    \label{Fig:trimer}
\end{figure}

We calculate the electronic structure and the vibrational properties (adiabatic and non-adiabatic) of Pt$_3$ and Pt$_5$ by performing fully relativistic calculations using version 6.4.1 of the \textsc{Quantum-Espresso} suite~\cite{0953-8984-21-39-395502,*Giannozzi_2017} and the compatible version of Thermo$\_$pw~\cite{PhysRevB.100.045115} for the non-collinear treatment of the magnetization densities.
We used version 3.3.0 of the fully relativistic ONCV pseudopotential \cite{PhysRevB.88.085117,*VANSETTEN201839} with Perdew-Burke-Ernzerhof exchange-correlation functional \cite{PhysRevLett.77.3865}
and a kinetic energy cutoff of 120~Ry.
A simple cubic Bravais lattice structure with a parameter of 10.6~\AA\ was used in order to minimize the interaction between the molecules and their copies.
The binding energy per atom of each cluster is obtained as $(nE_1-E_n)/n$ where $n$ is the number of atoms in the cluster and $E_1$ is the energy of the isolated atom.

We find that the lowest-energy structure of Pt$_3$ is an isosceles triangle with interatomic distances of 2.489~\AA\ and 2.501~\AA, as shown in Fig.~\ref{Fig:trimer}.
The binding energy per atom is $2.177$~eV. 
The HOMO-LUMO gap is $\Delta=137$~meV in agreement with \cite{PhysRevA.73.053201,YUAN20137}. 
The total magnetization is $1.58~\mu_B$.

For Pt$_5$ we obtain as the lowest energy structure a non-collinear magnetic trigonal bipyramid with the vertex atoms slightly shifted towards one side of the basis triangle.
The interatomic distances and the non-collinear magnetic atomic momenta are shown in Fig.~\ref{Fig:trimer}. 
The calculated binding energy per atom is $2.835$ eV in agreement with \cite{10.1063/1.3530799}.
The HOMO-LUMO gap is $\Delta=92$ meV and the total magnetization is $3.63~\mu_B$, in agreement with \cite{YUAN20137}. 
The smallness of $\Delta$ suggests the occurrence of large non-adiabatic effects both in Pt$_3$ and in Pt$_5$.

\begin{table*}[htp]
\setlength{\tabcolsep}{8pt} 
\renewcommand{\arraystretch}{1.1} 
\caption{Non-adiabatic effects in the optical modes of Pt$_3$ and Pt$_5$. From left to right, adiabatic mode index $\nu$, adiabatic vibrational frequencies $\omega_\nu$, non-adiabatic frequencies $\tilde\omega_\nu$, non-adiabatic frequencies $\tilde\omega_\nu^\prime$ obtained from the low-energy expansion of the dynamical matrix Eq.~\ref{Eq:force_const_Tay}, Cartesian components of the angular momentum $\bm\ell_\nu$ in units of $\hbar/2$.}    \label{Tab:frequencies}
\vspace{1mm}
\begin{tabular}{c S[table-format = 2.1] S[table-format = 3.1] S[table-format = 3.1] S[table-format = 3.1] c S[table-format = -1.3] S[table-format = -1.3] S[table-format = -1.3]}
 \hline \hline & \tabhead{$\nu$} &
 \tabhead{$\omega_\nu$ (cm$^{-1}$)} &
 \tabhead{$\tilde\omega_\nu$ (cm$^{-1}$)} & 
 \tabhead{$\tilde\omega_\nu^\prime$ (cm$^{-1}$)} & &
 \tabhead{$\ell_{\nu x}$ ($\hbar$/2)} &
 \tabhead{$\ell_{\nu y}$ ($\hbar$/2)} &
 \tabhead{$\ell_{\nu z}$ ($\hbar$/2)} \\
 \hline
 \multirow{3}{*}{Pt$_3$}
 & 1 & 102.4 & 100.6 & 102.5 & &-0.048 & 0.000 & 0.000 \\
 & 2 & 121.7 & 121.2 & 121.7 & & 0.000 & 0.000 & 0.000 \\
 & 3 & 217.7 & 217.7 & 217.7 & & 0.001 & 0.000 & 0.000 \\
 \hline
 \multirow{9}{*}{Pt$_5$}
 & 1 & 54.0 & 53.6 & 54.0 & & 0.000 & 0.000 & -0.064 \\
 & 2 & 71.1 & 71.1 & 71.1 & & 0.000 & 0.000 & -0.094 \\
 & 3 & 97.0 & 96.6 & 96.7 & & 0.001 & 0.000 & -0.089 \\
 & 4 & 103.3 & 103.5 & 103.5 & & 0.002 & 0.000 & -0.086 \\
 & 5 & 119.6 & 119.5 & 119.6 & & 0.002 & 0.000 & 0.000 \\
 & 6 & 134.8 & 134.7 & 134.7 & & -0.003 & 0.000 & 0.071 \\
 & 7 & 138.6 & 138.6 & 138.6 & & 0.000 & 0.000 & 0.093 \\
 & 8 & 169.4 & 169.4 & 169.5 & & 0.000 & 0.000 & 0.001 \\
 & 9 & 210.4 & 209.9 & 210.5 & & 0.000 & 0.000 & -0.003 \\
 \hline \hline
\end{tabular}
\end{table*}

Once the magnetic ground state is converged, we study the vibrational properties of the Pt clusters using linear response theory.
The adiabatic (static) optical frequencies of Pt$_3$ and Pt$_5$ are shown in the second column of Tab.~\ref{Tab:frequencies}.

Then the non-linear eigenvalue equation~\ref{Eq:secular_equation} is solved by simply evaluating the force-constant matrix $C_{\lambda\eta}(\omega)$ at different frequencies and by diagonalizing it. For each mode, the non-adiabatic (dynamical) vibrational frequency and polarization vectors can be found when the square root of the eigenvalue is equal to the value of the frequency fed into the dynamical force constant matrix.
The optical frequencies thus obtained are shown in the fourth column of Tab.~\ref{Tab:frequencies} for Pt$_3$ and Pt$_5$.
We find that both in Pt$_3$ and Pt$_5$, the non-adiabatic effects are small and do not modify the frequency of the optical modes substantially.
The vibrational frequencies $\tilde\omega_\nu^\prime$ obtained by solving the non-linear eigenvalue equation~\ref{Eq:secular_equation}, having replaced $C_{\lambda\eta}(\tilde\omega_\nu)$ with the low-energy expansion of the dynamical matrix Eq.~\ref{Eq:force_const_Tay}, are shown in column 5 of Tab.~\ref{Tab:frequencies}.

Non-adiabatic (dynamical) effects modify the oscillatory motion of the ions around their equilibrium positions.
The phonon ionic displacements are related to the polarization vectors through $\bm u_I=\text{Re}\left[\tilde{\bm{e}}_{I\nu} e^{-i\tilde\omega_\nu t}\right]$. 
In the adiabatic case the polarization vectors $\bm e_{I\nu}$ are real and the ionic motion reduces to a one-dimensional oscillation.
Instead, in the non-adiabatic case, the polarization vectors $\tilde{\bm e}_{I\nu}$ are complex and therefore the ions perform elliptical trajectories around their equilibrium positions.
Consequently, each ion gives rise to an orbital angular momentum perpendicular to the plane of the orbit.
For each mode, the angular momentum of the molecule is equal to the sum of the angular momenta of the rotating ions. It can be evaluated by replacing the non-adiabatic phonon polarization vectors $\tilde{\bm{e}}_{I\nu}$ into Eq.~\ref{Eq:mode_angular}.

\begin{figure}[htp]
\centering
\includegraphics[scale=1]{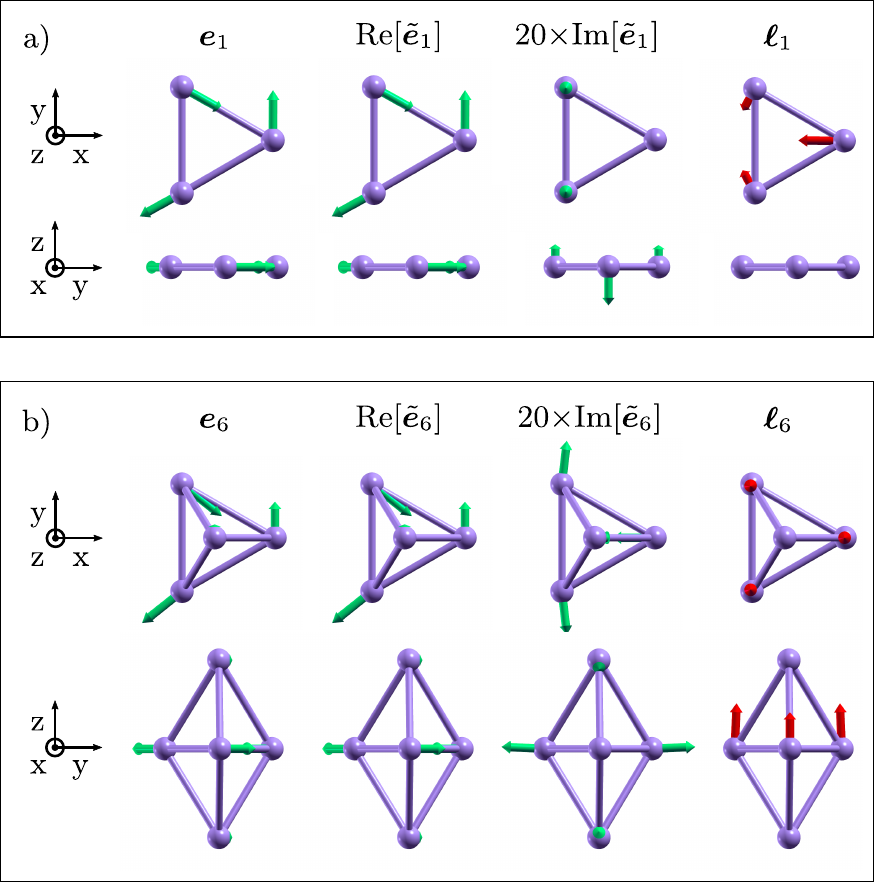}
\caption{From left to right, top (x-y) and side (y-z) representation of the adiabatic (static) polarization vectors $\bm e_\nu$, of the real and imaginary parts of the non-adiabatic (dynamical) polarization vectors $\tilde{\bm e}_\nu$ and of the vibrational angular momentum $\bm\ell_\nu$.
a) Asymmetric stretching mode of Pt$_3$ ($\nu=1$).
b) Asymmetric stretching mode of Pt$_5$ ($\nu=6$).}  \label{Fig:modes}
\end{figure}

As an illustrative example, we represent in Fig.~\ref{Fig:modes} the adiabatic and non-adiabatic polarization vectors of two stretching modes of Pt$_3$ and Pt$_5$.
In both cases the polarization vectors acquire an imaginary part and the non-adiabatic mode carry nonzero angular momentum.

The angular momentum of the optical modes of Pt$_3$ and Pt$_5$ is listed in the right hand side of Tab.~\ref{Tab:frequencies}.
Unexpectedly, we record a sizeable vibrational angular momentum even where the vibrational frequency is marginally altered by the non-adiabatic (dynamical) effects. The magnitude of these vibrational angular momenta
is of the same order of the typical values of the electron orbital momenta in itinerant ferromagnets \cite{Ceresoli}.

The total phonon angular momentum $\langle\bm L\rangle=\sum_\nu \bm\ell_\nu$ is nonzero because the angular momentum $\bm\ell_\nu$ is calculated at a different frequency for each mode $\nu$.
Since the angular momentum of the molecule must be conserved, a non-adiabatic variation of the electron angular momentum (spin plus orbital) must also occur in order to compensate the phonon contribution.
The calculation of such variation, however, requires
simulating the non-adiabatic dynamics of the whole molecule, which goes beyond the purpose of this work.

In conclusion, we have shown that in non-collinear magnetic molecules, non-adiabatic (dynamical) effects due to the electron-vibron coupling are time-reversal symmetry breaking interactions for the vibrational field. As in these systems the electronic wavefunction cannot be chosen as real, a non-zero geometric vector potential arises. As a result, an intrinsic non-zero phonon angular momentum occurs even for non-degenerate modes and in the absence of external time-reversal symmetry breaking probes. 
Our work provides the conceptual link between topology, electron-phonon interaction and the existence of a non-zero intrinsic phonon angular momentum in insulating non-collinear magnetic molecules.

As a proof of concept, we have demonstrated the magnitude of this topological effect by performing non-adiabatic first principles calculations on platinum clusters and by showing that vibrons display host sizeable intrinsic angular momenta with magnitude comparable to the typycal orbital electronic angular momenta in itinerant ferromagnets\cite{Ceresoli}.
As the same conclusions obtained for a molecule can be easily generalized to an insulating crystal, we expect that in any non-collinear magnetic system (solid or molecule) with strong electron-phonon interaction and sufficiently small gap non-adiabatic effects break time-reversal symmetry and generate sizeable intrinsic phonon angular momenta.

Finally, the question arises if the angular momenta of phonons can be observable in experiments. There are two cases in which it could be detected. The first is the case in which  a twofold degenerate mode at zone center occurs in the adiabatic phonon frequencies of the non-collinear magnetic system. As the time-reversal symmetry-breaking non-adiabatic term related to the Berry connection in Eq. \ref{Eq:Hamiltonian} and Eq. \ref{Eq:motion_freq} lowers the crystal symmetry, then the twofold degenerate mode could split in two different modes hosting different angular momenta. In this case, even if the angular momentum itself would not be observed, its effects on the phonon spectrum would. The second case is infrared absorption from left and right circularly polarized modes. As the vibrational angular momentum affects the atomic dipoles, the infrared intensities could be different for different circular polarizations.

We acknowledge IDRIS, CINES, TGCC, PRACE and CINECA for high performance computing resources.
We acknowledge P. Giannozzi and J. Carusotto for useful discussions.

\bibliography{bibliography}
\bibliographystyle{apsrev4-2}

\end{document}